\documentclass[%
 reprint,
 amsmath,amssymb,
 aps,
]{revtex4-2}

\usepackage{graphicx}
\usepackage{dcolumn}
\usepackage{bm}

\newcommand{\Kn}{{K\hspace{-.05cm}n}}

\begin{document}

\preprint{APS/123-QED}

\title{A correction tensor for approximating drag on slow-moving\\ particles of arbitrary shape}
\author{Duncan A. Lockerby}
\affiliation{%
University of Warwick, Coventry, CV4\,7AL, UK.
}%

\date{\today}

\begin{abstract}
A new form of the Cunningham correction factor is presented that requires no experimental fitting. It is expanded to provide a predictive heuristic for non-spherical particles, via definition of a ``correction tensor''. Its accuracy is tested against experiments and kinetic theory for the sphere, and stochastic solutions to the Boltzmann equation for a range of spheroids. It represents a simple, general tool for approximating transport properties of non-spherical micro/nano particles in a gas.  
\end{abstract}

                             
\maketitle
\noindent\textit{Introduction} -- Sub-micron pollutants can evade the body's respiratory defenses and deposit in the lung's alveoli, with exposure linked to stroke, heart disease, and cancer \citep{kwon_ultrafine_2020, donaldson_carbon_2006, poland_carbon_2008}. But particulate size is not the only issue --- mounting evidence suggests that particle \emph{shape} also plays a critical role in health outcomes \citep{thu_nanotoxicity_2023,donaldson_carbon_2006,li_work_2016,poland_carbon_2008}. This Letter proposes a simple and general method for predicting the transport properties of such slow-moving non-spherical sub-micron particles.

The scale of a particulate is characterized by the Knudsen number, $\Kn={\lambda}/{L}$, a dimensionless quantity representing the ratio of the gas's mean free path ($\lambda$) to the particle's characteristic length scale ($L$). For air at sea level, $\lambda\approx 65$ nanometers. (The Reynolds and Mach numbers are negligible for sub-micron particulates).

There is a wealth of experimental and theoretical data across the $\Kn$ scale for the drag on slowly translating spheres \cite{millikan_coefficients_1923,allen_re-evaluation_1982,allen_slip_1985,sone_external_1977,beresnev_motion_1990,kalempa_drag_2020} (which is related to their diffusivity via the Stokes-Einstein relation), but none, or very little, for the drag on non-spherical particles. Experiments are extremely challenging, and numerical simulations 
must involve solving the Boltzmann equation, or some approximation to it. Some progress in this has been made using the direct-simulation Monte Carlo (DSMC) method \citep{bird_molecular_1994}; however, DSMC is extremely computationally intensive, particularly for low-speed, low-$\Kn$, external flows \cite{bird_molecular_1994,livi_drag_2022, clercx_modeling_2024, tatsios_far-field_2025}.
Furthermore, studying non-spherical particles not only expands the parameter space but also complicates the analysis of resistance, as a single drag coefficient is insufficient. Instead, a ``resistance tensor'' must be determined to characterize motion.

Given these challenges, pragmatic means have been developed in the aerosol community to predict particle behavior across the $\Kn$ scale.
For example, there is a large body of work on predicting the mobility of aggregates of spheres at arbitrary $\Kn$ \cite{corson_analytical_2017,sorensen_mobility_2011}, as these can be a good model for many aerosol particles (e.g. soot).

A popular heuristic approach for spanning the $\Kn$ scale, applied to arbitrary geometries (not just agglomerates), is the Adjusted Sphere Method (ASM) proposed by Dahneke \cite{DahnekeASM1973} in 1973. The method's effectiveness has been demonstrated by both experiments \cite{cheng_drag_1988,rogak_measurement_1991,shapiro_characterization_2012,thajudeen_mobilities_2015} and DSMC \cite{livi_drag_2022,zhang_determination_2012}. The ASM predicts drag on an arbitrary particle by interpolating between the known ``continuum limit'' ($\Kn=0$) and the behavior in ``free-molecular flow'' ($\Kn\rightarrow\infty$). The main assumption being that, when scaled appropriately, the drag's transition between these limits, for any given particle, mirrors that of a simple sphere (for which reliable data exists).

Despite its success, ASM has significant drawbacks. The model is semi-empirical and based on results for a translating sphere, which makes its application to very non-spherical shapes and other resistance coefficients problematic. Furthermore, while the method accurately reproduces the asymptotic behavior for a non-spherical particle as $\Kn\!\rightarrow\!\infty$, it fails to do so as $\Kn\!\rightarrow 0$; it reproduces the correct limit value (the continuum one), but not the overall asymptotic behavior (discussed later). Finally, the method can be conceptually confusing; if used correctly, it requires defining a notional sphere and ``adjusted Knudsen number'' for each resistance coefficient. For a completely arbitrary geometry, with a full resistance tensor, this would result in a matrix of spheres and Knudsen numbers representing the same particle.

There is a need, then, for a simpler and more general heuristic approach for non-spherical particles.
\vspace{.2cm}

\noindent\textit{{The History of the Cunningham Correction Factor}}\,--\,In 1910, Ebenezer Cunningham published a short theoretical article on the settling velocity of spherical particles in a gas \cite{cunningham_e_velocity_1910}. The article's lasting legacy is an empirically-fitted expression that describes how drag on a spherical particle departs from Stokes law for non-zero $\Kn$.

To this day, the ``Cunningham correction factor'' is a fundamental tool in aerosol science and used in a vast array of associated technologies. For example, to understand the transport of atmospheric pollutants, for the design of filtration systems, in powder technologies, in lung deposition modeling, 
in contaminant control, in monitoring combustion emissions, and more.

For slow-moving spherical particles, the drag force ($D$) is linearly related to particle speed ($V$):
\begin{equation}\label{eq:linearFVRelation}
    D= \frac{6 \pi \mu R V }{C}
    \end{equation}
where $\mu$ is the dynamic viscosity, $R$ is the radius of the sphere, and $C$ is a correction factor, the form of which (normally attributed to Cunningham) is
\begin{equation}\label{eq:CCF}
   C_\mathrm{Cunningham}= 1+\Kn\,(\mathcal{A} + B e^{-c/\Kn})
\end{equation}
where $\mathcal{A}$, $B$ and $c$ are fitted coefficients and $\Kn=\lambda/R$.

 Although Cunningham is widely credited with (\ref{eq:CCF}), the exponential term doesn't feature at all in Cunningham's article. It was originally proposed by Knudsen and Weber \cite{knudsen_luftwiderstand_1911}, and only given a theoretical basis/interpretation in the 1920s by Millikan \cite{millikan_general_1923}.

Cunningham's correction was originally of a simpler form:
\begin{equation}\label{eq:COCF}
    C_\mathrm{original}= 1+\mathcal{B}\,\Kn \, ,
\end{equation}
where the coefficient $\mathcal{B}$ can be obtained theoretically. Cunningham's intention was to propose a heuristic expression capable of predicting particle settling velocity across the $\Kn$ scale, between the known limit in continuum conditions (i.e. Stokes drag; $C=1$) and the correct asymptotic behavior in the limit of free-molecular flow:
\begin{equation}\label{eq:hlim}
   C \sim  \mathcal{B}\, \Kn  \hbox {\quad as \quad  } \Kn \rightarrow\infty  \, .
\end{equation}

Millikan \cite{millikan_general_1923}, over 10 years later, pointed out a failing in Cunningham's expression (\ref{eq:COCF}), in that it did not reproduce the correct asymptotic behavior at small Knudsen numbers ($   C \nsim  (1+ \mathcal{B}\, \Kn)  \hbox { as } \Kn \rightarrow 0 $).
It is correct in form, but the coefficient, which can be extracted from the analytical work of Basset \cite{basset_treatise_1888} on slip flow, is not equal to $\mathcal{B}$. Instead,
\begin{equation}\label{eq:llim}
   C \sim  (1+\mathcal{A}\, \Kn) \hbox {\quad as \quad  } \Kn \rightarrow 0 \, .
\end{equation}
For this reason, Millikan \cite{millikan_general_1923} proposed a pragmatic modification that would allow both asymptotic behaviors to be captured. Millikan modified the prefactor to $\Kn$ in Cunningham's expression (\ref{eq:COCF}) so that it would smoothly transition from $\mathcal{A}$ to $\mathcal{B}$ in the appropriate limits:
\begin{equation} \label{eq:CCFv2}
    C_\mathrm{Cunningham} =1+\Kn\,(\mathcal{A}+(\mathcal{B}-\mathcal{A})e^{-c/\Kn})\, ,
\end{equation}
where $c$ is a parameter that Millikan described as the ``rapidity of shift'', which must be fitted to experimental data.
Note, (2) and (6) are equivalent to each other; $B=\mathcal{B}-\mathcal{A}$. 

It was Millikan's intervention then, that changed the course of history for the Cunningham correction factor: from its origin as a predictive heuristic to its modern use as a fitting function \cite{c_n_davies_definitive_1945, allen_re-evaluation_1982,allen_slip_1985}; and from its original form (\ref{eq:COCF}) to the one actually due to \citet{knudsen_luftwiderstand_1911}. Millikan's analysis was partly an attempt to claim some intellectual ownership of Knudsen and Weber's equation by providing a physical basis for it. It is ironic, then, that neither Millikan, Knudsen nor Weber typically gets the credit.
\vspace{.2cm}

\noindent\textit{{A New Correction Factor}} -- Millikan's sound argument in 1923 was that the Cunningham heuristic needed to be modified so that the analytical limit of (\ref{eq:llim}) could also be captured. However, for this purpose, his introduction of two new parameters into Cunningham's expression (to satisfy one more condition), is needlessly complicated. It is also less general, of course, because the spare coefficient $c$ needs to be fitted to experimental data.

Millikan did not spot (or at least did not mention) that a simpler expression exists that satisfies both (\ref{eq:hlim}) and (\ref{eq:llim}), without need for a fitting parameter:

\begin{equation}\label{eq:NewCC}
    C_{\mathrm{new}}=e^{-(\mathcal{B}-\mathcal{A})\Kn}+\mathcal{B}\Kn \, .
\end{equation}

\noindent\textit{Test Case: Spherical Particles} --- The new form of correction factor (\ref{eq:NewCC}), being free of empirical parameters, can serve as a powerful predictive heuristic. Although the focus of this Letter is non-spherical particles, a sphere is the logical first benchmark for evaluating its accuracy, as reliable data is readily available.

For a sphere, the coefficient $\mathcal{A}$ can be obtained by solving the Stokes equations with a Maxwell slip boundary condition \cite{maxwell_stresses_1879} (as done by Basset \cite{basset_treatise_1888}), with the result taken to first order in $\Kn$ (as done by Millikan \cite{millikan_general_1923}). In this way it is easy to show, for the case of the sphere, that $\mathcal{A}=\beta$ where $\beta$ is the boundary condition's `slip' coefficient. Note, Maxwell derived his slip boundary condition for planar surfaces, but it's applicable to any particle in the limit of $\Kn\rightarrow0$ (the limit at which $\mathcal{A}$ is evaluated). For diffuse molecular reflection, Maxwell found $\beta=1$, but modern kinetic theory predicts a slightly higher value, typically in the range $\beta=1.11$--$1.15$, depending on the specific molecular-collision model \cite{sharipov_data_2004}. Young proposed a working average of $\beta=1.13$ \cite{young_thermophoresis_2011}, which will be adopted here and for the rest of this Letter. The coefficient $\mathcal{B}$ comes directly from Epstein's free-molecular result \cite{epstein_resistance_1924}. For diffuse reflection of gas molecules at the sphere surface, $\mathcal{B}=18/(8+\pi)$.

Figure\,1 compares the prediction of equation (\ref{eq:NewCC}) with the data from Millikan \cite{millikan_general_1923}, relatively recent experiments \cite{allen_slip_1985}, and modern kinetic theory \cite{beresnev_motion_1990}. Given equation (7) is heuristic, with no fitting parameters, its prediction is exceptionally good.
\begin{figure}[h]
\includegraphics[width=1\columnwidth]{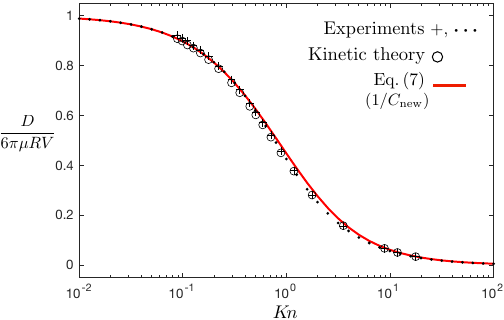}
\caption{Drag on a slowly translating sphere against $\Kn$. Comparison of Millikan's data \cite{millikan_general_1923} ($+$), experiments of \citet{allen_slip_1985} (\,$\boldsymbol{\cdots}$), kinetic theory of \citet{beresnev_motion_1990} (\raisebox{0.2ex}{\scalebox{0.6}{$\bigcirc$}}) and the proposed heuristic ($1/C_{\mathrm{new}}$, ---\,), equation\,(\ref{eq:NewCC}).} 
\end{figure}
\newpage
\noindent\textit{Application to Non-spherical Particles} -- The forms of the asymptotic limits (\ref{eq:hlim}) and (\ref{eq:llim}) are the same for particles of arbitrary shape. 
In the continuum limit ($\Kn\rightarrow0$), a small departure from Stokes law can be expressed as a series expansion truncated to first order in $\Kn$; a correction factor in the form of ($1+ \mathcal{A}\Kn$) follows.
In the free-molecular limit, there are no gas-molecule interactions (i.e.~no collisions): the effect of a gas molecule on the resistance to particle motion is independent of any other. Given the low particle speed (relative to the average gas molecule), the resistance generated by the gas (for a given particle motion) as $\Kn\rightarrow\infty$ becomes proportional to the density of the gas, and the correction factor proportional to $\Kn$; as in (\ref{eq:hlim}). 

The core idea of this Letter is: if $\mathcal{A}$ and $\mathcal{B}$ can be found for a particle, equation (7) can be used to approximate the gas's resistance to its motion across the entire $\Kn$ scale --- whatever the particle shape.
\vspace{.2cm}

\noindent\textit{A Correction Tensor} --
When a particle moves slowly enough that the gas flow responds linearly to its motion, the resistance to the particle's translation is most generally described by a resistance tensor \cite{happel_low_1983}:
\begin{equation}\label{eq:Kv}
    \bm{f}=-\bm{K}  \boldsymbol{\cdot}  \bm{v}
\end{equation}
 where $\bm{f}$ is the force on the particle, $\bm{v}$ is the particle velocity (relative to the gas), and $\bm{K}$ is a 3$\times$3 tensor, whose components depend (potentially independently) on $\Kn$. 
 It is relatively easy to calculate the resistance tensor in certain limits:
\begin{equation}
    \bm{K}^{0}, \bm{K}^{1} = \lim_{\Kn \to 0}\left(\bm{K}, \frac{\mathrm{d}\bm{K}}{\mathrm{d}\Kn}\right),
    \quad \bm{K}^{\infty} = \lim_{\Kn \to \infty}(\Kn \bm{K}).
\end{equation}
From these, following the spirit of Cunningham, but using equation (\ref{eq:NewCC}), it is possible to propose a simple element-wise correction to the continuum resistance tensor:
\begin{equation}\label{eq:KC}
K_{ij}=K_{ij}^{\mathrm{0}}/C_{ij}
\end{equation}
where 
$C_{ij}=e^{-(\mathcal{B}_{ij}-\mathcal{A}_{ij})\Kn}+\mathcal{B}_{ij}\Kn$, and where the correct asymptotic behavior at both limits is ensured by setting ${\mathcal{A}_{ij}} =-K^{\mathrm{1}}_{ij}/K^{\mathrm{0}}_{ij}$  and ${\mathcal{B}_{ij}} = K^{\mathrm{0}}_{ij}/K^{\mathrm{\infty}}_{ij} $.
\vspace{0.2cm}

\noindent\textit{Evaluating the Correction Tensor} --- To evaluate $\bm{C}$ for a given particle, all that is required is $\bm{K}^{\mathrm{0}}$, $\bm{K}^{\mathrm{1}}$ and $\bm{K}^{\mathrm{\infty}}$. The continuum resistance tensor $\bm{K}^{\mathrm{0}}$ can be determined by solving the Stokes equations with a no-slip boundary condition; for complex geometries this can be done, for example, using the boundary element method \cite{pozrikidis_boundary_1992} or the Method of Fundamental Solutions (MFS) \cite{jordan_method_2025}. For some geometries, analytical results exist (for example, the spheroid \cite{oberbeck_ueber_1876,happel_low_1983}; see the appendix).

The free-molecular resistance tensor, $\bm{K}^{\mathrm{\infty}}$, for arbitrary convex particles, can be obtained by evaluating the following integral over the particle surface, $S$ \cite{DahnekeFM1973, halbritter_torque_1974}:
\begin{equation}\label{eq:Kinf}
    \bm{K}^{\infty} = \frac{\mu}{L}   \int_{S}  \left( \frac{\sigma}{2} \bm{I} + \gamma\, \bm{n}\bm{n}\right)\, dS 
    \end{equation} 
where $\bm{I}$ is the identity tensor, $\bm{n}$ is an outward-facing surface normal, $\gamma=\left(8+ \pi\sigma-6\sigma\right)/4$ and $\sigma$ is the accommodation coefficient (typically $\sigma\approx1$).  The analytical result for spheroids \cite{DahnekeFM1973} is included in the appendix. For more complex convex particles, the surface integral (\ref{eq:Kinf}) can be performed numerically (e.g., using the MFS \cite{lockerby_integration_2022}). For non-convex particles, DSMC can be adopted \cite{gallis_approach_2001,chinnappan_transport_2019} which is very efficient in free-molecular flow.

The first-order resistance tensor, $\bm{K}^{\mathrm{1}}$, is not so straightforward.
For a sphere, a simple analytical solution to the Stokes equations with slip boundary conditions exists \cite{basset_treatise_1888}. This can be truncated to first order in $\Kn$, as Millikan did \cite{millikan_coefficients_1923}. However, for non-spherical particles, such solutions generally do not exist. Even the spheroid slip solution requires an ``infinite-series form of semi-separation of variables'' \cite{keh_slow_2008}.

Fortunately, a very convenient technique has been developed, relatively recently, for evaluating $\bm{K}^{\mathrm{1}}$ directly \cite{ramachandran_dynamics_2009,stone_interfaces_2010}. It exploits the reciprocal theorem \cite{happel_low_1983}, and has been employed to derive expressions for the first-order slip/$\Kn$ effect on drag around spheroids \citep{masoud_reciprocal_2019}, Janus spheres \citep{ramachandran_dynamics_2009}; and a range of other configurations \cite{lockerby_first-order_2025}. Adapting the form in \cite{lockerby_first-order_2025} we can write:
\begin{equation}\label{eq:K1}
\bm{K}^{\mathrm{1}}=K_{ij}^{\mathrm{1}}=-\frac{L}{\mu} \int_{S} \bm{\tau^{i}}\boldsymbol{\cdot}\bm{\tau^{j}}\, \, dS ,
\end{equation}
where $\bm{\tau^i}$ is the surface shear-stress vector generated by a unit particle velocity in the $i^\mathrm{th}$ direction, from a \emph{no-slip} solution to the Stokes equations. Masoud and Stone's closed-form expression for the spheroid  \cite{masoud_reciprocal_2019} is provided in the appendix.
\vspace{.3cm}

\noindent\textit{Results} --
 In the absence of comprehensive experimental data for drag on slow-moving non-spherical particles across the $\Kn$ scale, DSMC simulations (stochastic solutions to the Boltzmann equation) represent the accepted and most reliable benchmark. Besides DSMC simulations for aggregations of spheres \cite{zhang_determination_2012}, only spheroids have been properly studied  \cite{livi_drag_2022,clercx_modeling_2024,tatsios_far-field_2025,zhang_evaluation_2025}; the most careful and detailed being by \citet{livi_drag_2022} and \citet{clercx_modeling_2024}.  
 
Figure 2 compares data from \citet{clercx_modeling_2024} to equation (\ref{eq:KC}), for drag on various aspect ratio, prolate and oblate spheroids, as a function of $\Kn$. Here the spheroid is defined by $(x/a)^2 + (y/b)^2 + (z/b)^2 = 1$, where $x$ is along its axis of revolution; for all cases $L=\sqrt[3]{ab^2}$. The general level of agreement in Figure 2 is excellent.
\begin{figure}[ht]
\includegraphics[width=1\columnwidth]{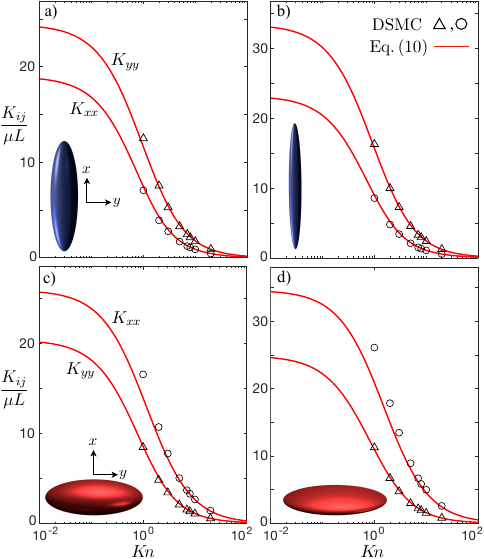}
\caption{\label{fig:2} Resistance tensor components for prolate (a,\,b) and oblate (c,\,d) spheroids of aspect ratio 4 (a,\,c) and 10 (b,\,d). Motion parallel ($\triangle$, $K_{x\hspace{-.01cm}x}$) and perpendicular (\raisebox{0.3ex}{\scalebox{0.6}{$\bigcirc$}}, $K_{y\hspace{-.01cm}y}$) to the polar axis. Comparison of DSMC \cite{clercx_modeling_2024} to Eq.\,(\ref{eq:KC}).}
\end{figure}
\begin{figure}[!ht]
\vspace{1cm}
\includegraphics[width=1\columnwidth]{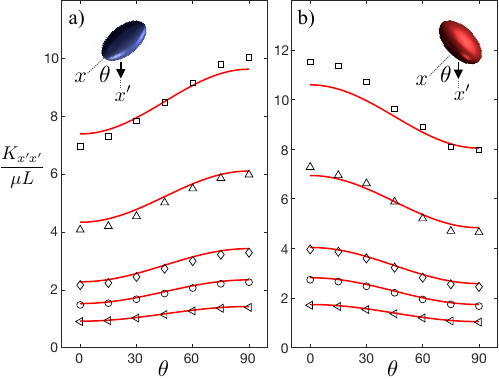}
\caption{Resistance component in the direction of motion ($x'$) for prolate (a) and oblate (b) spheroids of aspect ratio 2, against $\theta$, the angle (in degrees) between $x'$ and $x$ (where $x$ is the spheroid's axis of revolution). DSMC data \cite{livi_drag_2022} at $\Kn=1(\square),\,5(\triangle),\,7(\Diamond),\,9($\raisebox{0.3ex}{\scalebox{0.6}{$\bigcirc$}}$)\,$and $10(\triangleleft);$ Eq.\,(10) (---).}
\end{figure}

Figure 3 compares (\ref{eq:KC}) with the data of \citet{livi_drag_2022} for aspect ratio 2 spheroids, at various $\Kn$, as a function of orientation. The benefit of the resistance tensor description (\ref{eq:Kv}), is that the drag for any particle orientation can be determined by a simple transformation: $\bm{K}'=\bm{R}\cdot \bm{K} \cdot \bm{R}^\top$, where $\bm{R}$ is a rotation matrix. For Figure 3, the flow direction ($x'$) is rotated from the spheroid's axis of revolution ($x$) by a single-axis rotation $\theta$ about $z$: $\bm{R}=\bm{R}_z(\theta)$. Again, the general level of agreement is very good, with the greatest discrepancy at $\Kn=1$, where the heuristic is farthest from either limit and DSMC is hardest to perform accurately.

\noindent\textit{{Discussion and Future Work}} --
The results in Figures 1-3 demonstrate that the heuristic proposed has considerable predictive power. Importantly, there are no fitting parameters: the predictions are all based on known analytical results (the ones for a spheroid are provided in the appendix, for convenience). However, it is also important to stress that the prediction is based on an interpolation across the highly complex ``transition regime'' ($0.1<\Kn<10$). It is only an approximation, and no replacement for future, much-needed, experiments and Boltzmann-equation solutions for non-spherical particles. 

The predictions have not been tested in the low transition regime, due to the compounded difficulties in performing DSMC there \cite{tatsios_far-field_2025}. It is fair to point out that the Adjusted Sphere Method (ASM) also performs well in the high transition regime. However, a disadvantage of the ASM, other than its conceptual issues and need for empirical input, is that it can't ensure $\lim_{\Kn\rightarrow0}(\mathrm{d} \bm{K}/\mathrm{d}\Kn)=\bm{K}^\mathrm{1}$, which equation (\ref{eq:KC}) does by design. For an oblate spheroid of aspect ratio 3, the ASM underpredicts the steepness of the $K_{xx}(\Kn)$  curve at $\Kn=0$ by about 20\%; this increases to nearly 30\% for aspect ratio 10. It is also fair to point out here that \citet{livi_drag_2022} and \citet{clercx_modeling_2024}  also proposed useful interpolation schemes through their data, but these used spheroid DSMC data for fitting, and are thus only applicable to the spheroid.

The correction factor (\ref{eq:NewCC}) assumes that $\mathcal{B}-\mathcal{A}>0$. This is the case for all the results presented here, but not true in general. In the event that $\mathcal{B}<\mathcal{A}$, a practical fix is to set the exponent ($\mathcal{B}-\mathcal{A}$) to zero. This will sacrifice accurate asymptotic behavior as $\Kn\rightarrow0$ for a sensible correction across the $\Kn$ scale.

Brenner proved that the continuum resistance tensor ($\bm{K}^\mathrm{0}$) is symmetric \cite{happel_low_1983}. Quick inspection of (\ref{eq:Kinf}) and (\ref{eq:K1}) reveals that $\bm{K}^\mathrm{1}$ and $\bm{K}^\mathrm{\infty}$ are also symmetric; it follows, then, that $\bm{C}$, and any predicted resistance tensor, will also be symmetric. 
As discussed in \cite{happel_low_1983}, an alternative proof exists for the symmetry of $\bm{K}^\mathrm{0}$ due to Landau and Lifshitz \cite{landau_statistical_1958,landau_fluid_1959} that doesn't require invocation of the fluid equations at all, and is purely based on thermodynamic considerations. It thus seems likely that the resistance tensor retains symmetry across the $\Kn$ scale as predicted by (\ref{eq:KC}). One of the consequences of this is that the ``sine-squared drag law'' should hold at all $\Kn$, which is also supported by DSMC \cite{livi_drag_2022,clercx_modeling_2024,zhang_evaluation_2025,bird_molecular_1994}.

The form of the correction tensor tacitly assumes that the direction of the principal axes remains constant across the entire $\Kn$ scale. This assumption may not hold true, and the extent of its validity requires further investigation. Future work should also include applying the heuristic correction to the full 6$\times$6 resistance tensor for the study of particles with rotational-translational coupling \cite{brenner_coupling_1965}.
\vspace{0.2cm}

\noindent\textit{Data availability} --
A short script for evaluating the spheroid resistance tensor as a function of $\Kn$, used in Figures 2 and 3, will be made available on publication.

\bibliographystyle{apsrev4-2}
\bibliography{references}
\newpage
\appendix*
\begin{widetext}
\section{$\bm{K}^{\mathrm{0}}$, $\bm{K}^{\mathrm{1}}$ and $\bm{K}^{\mathrm{\infty}}$ for spheroids}
\noindent The continuum \cite{happel_low_1983}, first-order \cite{masoud_reciprocal_2019}, and free-molecular \cite{DahnekeFM1973} resistance tensors for a spheroid defined by $(x/a)^2 + (y/b)^2 + (z/b)^2 = 1$ are
\begin{align*}
&\bm{K}^{\mathrm{0}} =16\,\pi \mu  \,b \,e^3 \left(
\frac{\bm{i}_x \bm{i}_x}{2 e\, r+\left(4 e^2-2\right) \sin ^{-1}(e)} + \frac{\bm{i}_y\bm{i}_y+\bm{i}_z \bm{i}_z}{(2\,e^2 + 1)\,\mathrm{sin}^{-1}(e) -e\,r}
\right) \, , \\
&\bm{K}^{\mathrm{1}}=-32\,\pi \mu L \,e^3 \beta \left(\bm{i}_x \bm{i}_x \frac{e-r^2 \,\mathrm{tanh}^{-1}(e)}{2\,{{\left(e\,r+\mathrm{sin}^{-1}(e)\,{\left(2\,e^2 -1\right)}\right)}}^2 }
 +( \bm{i}_y\bm{i}_y+\bm{i}_z \bm{i}_z) \frac{\mathrm{tanh}^{-1}(e)\,{\left(e^2 +1\right)-e}}{{{\left(e\,r-\mathrm{sin}^{-1}(e)\,{\left(2\,e^2 +1\right)}\right)}}^2 }\right)  \hbox{ and} \\
&\bm{K}^{\mathrm{\infty}} =\frac{\mu}{L} \frac{\pi b^2}{e^3} \biggl(\bm{i}_x \bm{i}_x \left[ \sigma e^3 + 2e \gamma + r^2 (\sigma e^2 - 2\gamma) \,\mathrm{tanh}^{-1}(e)\right]\nonumber \\ & \hspace{2cm}+(\bm{i}_y\bm{i}_y+\bm{i}_z \bm{i}_z)\Bigl[-\gamma e r^2 + e^3 \sigma    + r^2 (\gamma(1 + e^2) + e^2 \sigma) \mathrm{tanh}^{-1}(e)\Bigr] \biggr)\, ,
\end{align*}
where $\bm{i}_{x,y,z}$ are unit vectors, $r=a/b$, and $e=\sqrt{1-r^2}$ (which is real for oblate spheroids and imaginary for prolate).

\end{widetext}

\end{document}